# Thermally stimulated luminescence of oxygen-deficient zirconia nanotubes

A.S. Vokhmintsev, I.A. Petrenyov, R.V. Kamalov,
M.S. Karabanalov, I.A. Weinstein

NANOTECH Centre, Ural Federal University,
Mira str., 19, Ekaterinburg, Russia, 620002

a.s.@vokhmintsev; i.a.weinstein@urfu.ru

$ZrO_2$ nanotubular arrays with intrinsic defects are a promising solid-state basis for the development of devices for detecting, storing, and converting energy. Layers of the self-ordered zirconia nanotubes of 5 μm length and 30 nm diameter, containing oxygen vacancies and their complexes, have been synthesized by anodic oxidation. The spectrally resolved TSL (thermally stimulated luminescence) above room temperature for the samples exposed by UV-irradiation with an energy of 4.1 eV have been studied. Two emission bands with maxima near 2.5 and 2.8 eV, associated with radiative relaxation of T and $F^+$ centers, respectively, have been found. An analysis of the measured glow curves within the framework of general order kinetics established the presence of four TSL peaks caused by charge carriers traps with activation energies of 0.7–0.8 eV. A band diagram is proposed to explain the role of optically active centers based on electron (vacancies in positions of three- and four-coordinated oxygen) and hole (interstitial oxygen ions) traps in observed TSL processes during the irradiation and the followed heating of anion-deficient $ZrO_2$ nanotubes.

# 1. Introduction

To date, various high technology industries are extensively engaged in exploiting multifunctional structures based on oxide nanoporous and nanotubular high-specific-surface-area arrays [1–3]. In particular, due to the high chemical, mechanical, radiation, and temperature stability [3–5] zirconia nanotubes are used to create new batteries, capacitors, fuel cells, catalysts, memristor cells, dosimetric detector media, etc. [6–9].

It is known that anion- and cation-based point lattice imperfections – F-, $F^+$- and $Zr^{3+}$-centers aggregated by T-defects [10], vacancy clusters [11], etc. – are optically active centers to form the features of the luminescent response of zirconia to various external impacts. Previously, photoluminescence spectroscopy (PL) [12–19] and galvanoluminescence [20] techniques have provided for studying $ZrO_2$ nanotubes grown by anodization. In addition, the bandgap width of $E_g =$ 5.66 eV has been determined by the diffuse scattering method for nanotubular arrays of $ZrO_2$ annealed in air at 400 °C and consisting of a mixture of tetragonal and monoclinic phases. [21]. In addition, several papers report on investigation of $ZrO_2$ powders with initially monoclinic symmetry, which were heat-treated further in air [9,22–27], in vacuum [28] and in an argon atmosphere [29] using thermally stimulated luminescence (TSL) in the range of 400–650 nm. For samples exposed to an electron beam [9,22,28], β- [23–26] and UV-radiation [27,29], the presence of active traps of charge carriers was experimentally confirmed by an emission in the ranges of below room temperature (223–293K) [29] and (300–750K) [9, 22–28]. Currently, we are not aware of published works on the study of $ZrO_2$ nanotube structures using the thermally stimulated luminescence (TSL) method. The present work aims to explore the mechanisms of TSL processes that occur in $ZrO_2$ nanotubes after UV irradiation.

# 2. Samples and Methods

Nanotubular samples of zirconium dioxide were synthesized by anodizing Zr foil with an Hf < 1% impurity [30]. 100 μm thick Zr-foil underwent the following treatment: preliminary degreasing, immersion in an ultrasonic bath, acidulation by solution $HF:HNO_3:H_2O$ = 1:6:20, washing-up with distilled water, and drying in air. Anodizing was conducted in a two-electrode electrochemical cell at a constant voltage of 20 V and an anode temperature of 10 °C for 6 h. An ethylene glycol solution containing 5 wt. % $H_2O$ and 1 wt. % $NH_4F$ served as an electrolyte. The above-mentioned procedure is described in detail in [30].

The morphological parameters of the synthesized arrays were studied using a scanning electron microscope (SEM) AURIGA, Carl Zeiss and transmission electron microscope (TEM) JEM-2100, JEOL. The image shown in Figure 1 contains an oxide layer consisting of ≈30 nm external diameter self-ordered nanotubes. When synthesized under the above conditions, the $ZrO_2$ arrays have the thickness (length of nanotubes) of about 5 μm [30].

After synthesis, the $ZrO_2$ layers were heated to 800 K in air to remove electrolyte residues. The final samples for taking TSL measurements involved a rectangular-shaped zirconium substrate of 12×7 mm in size, one-side-covered by a layer of $ZrO_2$ nanotubes.

The spectrally resolved TSL response was recorded on a Perkin Elmer LS 55 luminescent spectrometer equipped with a unique high-temperature attachment [31, 32] in the range $T$ = 300–773 K at a heating rate of $\beta$ = 2 K/s, in the range from 390 to 590 nm with a step of 10 nm. The optical gap in the recording path amounted to 20 nm. Before irradiation, the samples were preliminarily heated to 773 K followed by cooling to room temperature inside the dark chamber of the spectrometer. Irradiation was carried out with monochromatic UV radiation with a wavelength of $\lambda$ = 300 nm (4.1 eV) at room temperature. The dose (energy density) of UV radiation was 1 mJ/cm². The experimental TSL curves were analyzed within the formalism of general order kinetic processes [33]:

$$I(T) = s" \cdot n_0 \cdot \exp\left(-\frac{E_a}{kT}\right) \cdot \left[1 + \frac{s" \cdot (b-1)}{\beta} \cdot \int_{T_0}^{T} \exp\left(-\frac{E_a}{kT'}\right) \cdot dT'\right]^{-\frac{b}{b-1}}, \quad (1)$$

where $I(T)$ is the TSL intensity; $s"$ is an effective frequency factor, s$^{-1}$; $n_0$ is the initial concentration of filled traps after irradiation, m$^{-3}$; $E_a$ is the trap activation energy, eV; $k$ is the Boltzmann constant, eV/K; $T$ is temperature, K; $b$ is the order of kinetics; $\beta$ is the heating rate, K/s; $T_0 = 300$ K is the initial temperature, K.

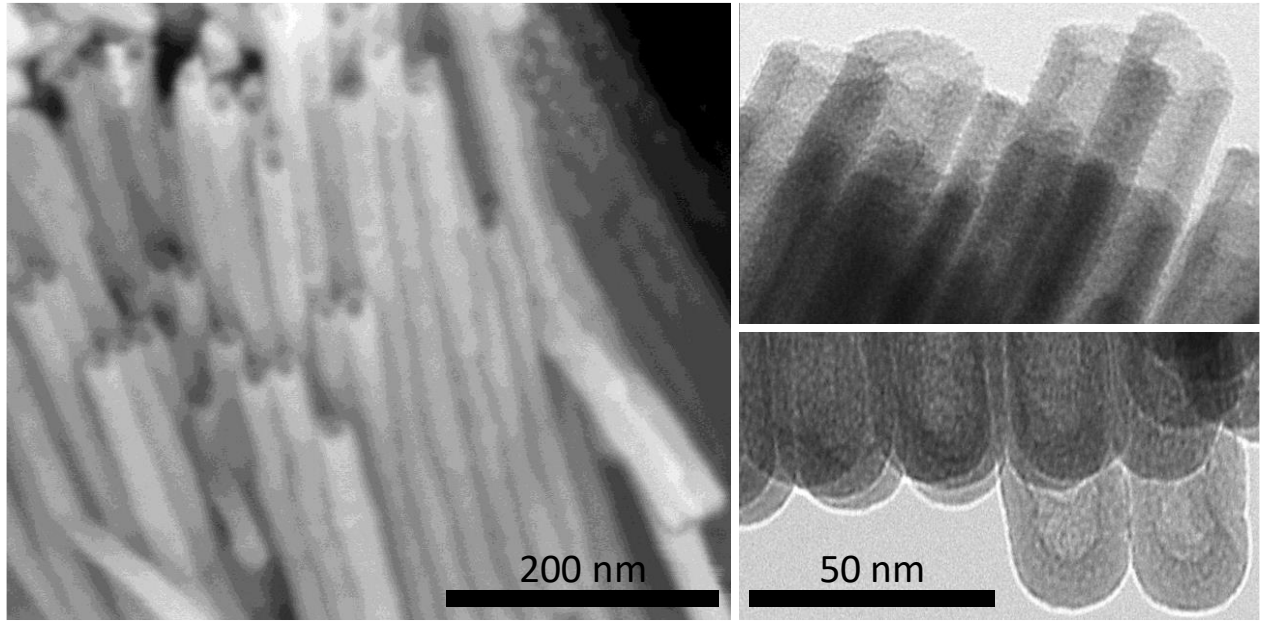

**Figure 1.** SEM (left) and TEM (right) images of synthesized ZrO$_2$ nanotubes.

The energy and kinetic characteristics of trapping centers were determined by approximating the shape of the TSL curves by the superposition of several independent peaks of the general order of kinetics according to Eq. (1). This technique is successfully used in TSL dosimetry of ionizing radiation in studying wide-gap materials [23–26, 34–36].

## 3. Experimental Results

Figure 2 displays the 3D plot of the TSL data measured in the emission bands investigated. It can be seen a wide temperature peak with maximum at ≈ 340 K. Besides, an extended, less intense shoulder being observed in the region of 400–550 K. The highest intensity of the TSL response is characteristic of the 490 nm band. Let us analyze the resulting luminescence spectrum to elucidate the possible nature of the recombination centers that form the recorded TSL emission.

## 4. Discussion

*4.1. Spectral Characteristics of TSL*

Figure 3 shows the TSL emission spectrum at 340 K in ZrO$_2$ nanotubes on the photon energy. Each point of the spectrum corresponds to the area beneath the experimental TSL curve (Fig. 2) in the corresponding temperature range.

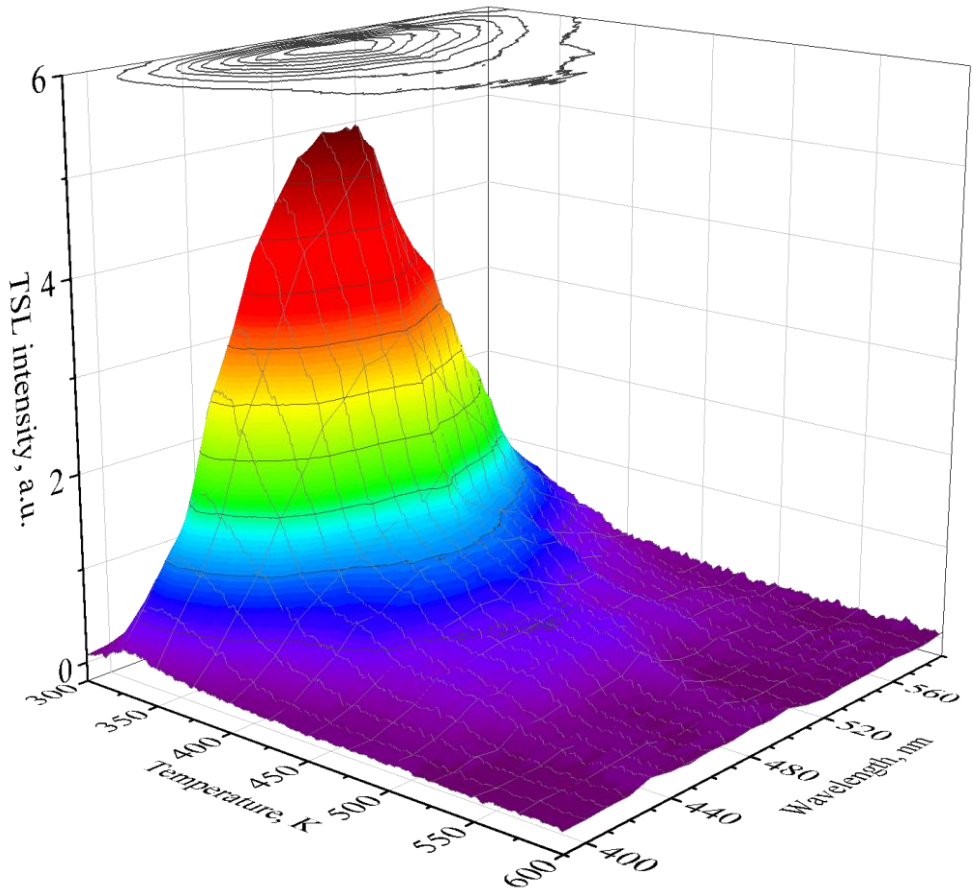

**Figure 2.** 3D plot of the experimental TSL response measured at different wavelengths.

Obviously, the resulting TSL spectrum can be described by two Gaussian components G1 and G2 with maximum energies $E_{max}$ = 2.51 ± 0.06 and 2.81 ± 0.09 eV and half-widths $\omega$ = 0.45 ± 0.07 and 0.39 ± 0.09 eV, respectively. The coefficients of determination $R^2$ > 0.99. The results obtained are consistent with our findings concerning the PL properties of $ZrO_2$ nanotubular layers [12], for which two Gaussian luminescence bands are also recorded with $E_{max}$ = 2.43 ± 0.01 and 2.68 ± 0.03 eV, as well as $\omega$ = 0.35 ± 0.01 and 0.48 ± 0.02 eV, respectively. Thus, it can be concluded that two luminescent centers are involved in the thermal activation processes under study. In addition, when studying monoclinic $ZrO_2$ powders, the TSL [9, 22–29], PL [9,26,28,29], and pulsed cathodoluminescence [9,28] spectra showed a broad nonelementary luminescence band at a maximum of 480–500 nm (2.58–2.48 eV).

Thus, the close spectral composition of the luminescence for different types of excitations in $ZrO_2$ structures with different morphology may indicate the predominance of the monoclinic phase in the samples. This fact agrees with X-ray diffraction studies of $ZrO_2$ nanotubes that are characterized by the formation of a monoclinic phase as a result of annealing at 400 °C for 1 h in air [21].

The emission in the G1 band (≈2.4 eV) may evidence the recombination between holes of the valence band and electrons captured by $Zr^{3+}$ ions [10]. There is currently no consensus on the nature of the G2 band (2.5–2.7 eV) [28]. On the one hand, this luminescence is presumably caused by impurity relaxation in $ZrO_2$ with a very low $Ti^{3+}$ concentration [37]. In our case, initially, the foil does not have titanium ions and the synthesis excludes the latter's appearance in the samples. On the other hand, the luminescence at 2.5 eV appears to be attributed to the processes of radiative relaxation of oxygen vacancies in the bulk or on the sample surface [38,39]. In this context, an imortant role may be played by complex defects, including associated oxygen vacancies [40] or T-centers based on the bound state between two neighboring oxygen vacancies positioned in [111] direction, with the $Zr^{3+}$ ion [10].

When possibly participating in the mechanisms of the recorded TSL, the T-centers are in good agreement with their position on the excited level that is 4.2 eV above the top of the valence band [10]. It is worth noting that the method used for the synthesis of the ZrO nanotubular samples and their developed surface contribute to the accumulation of a large number of oxygen vacancies and $Zr^{3+}$ ions, respectively.

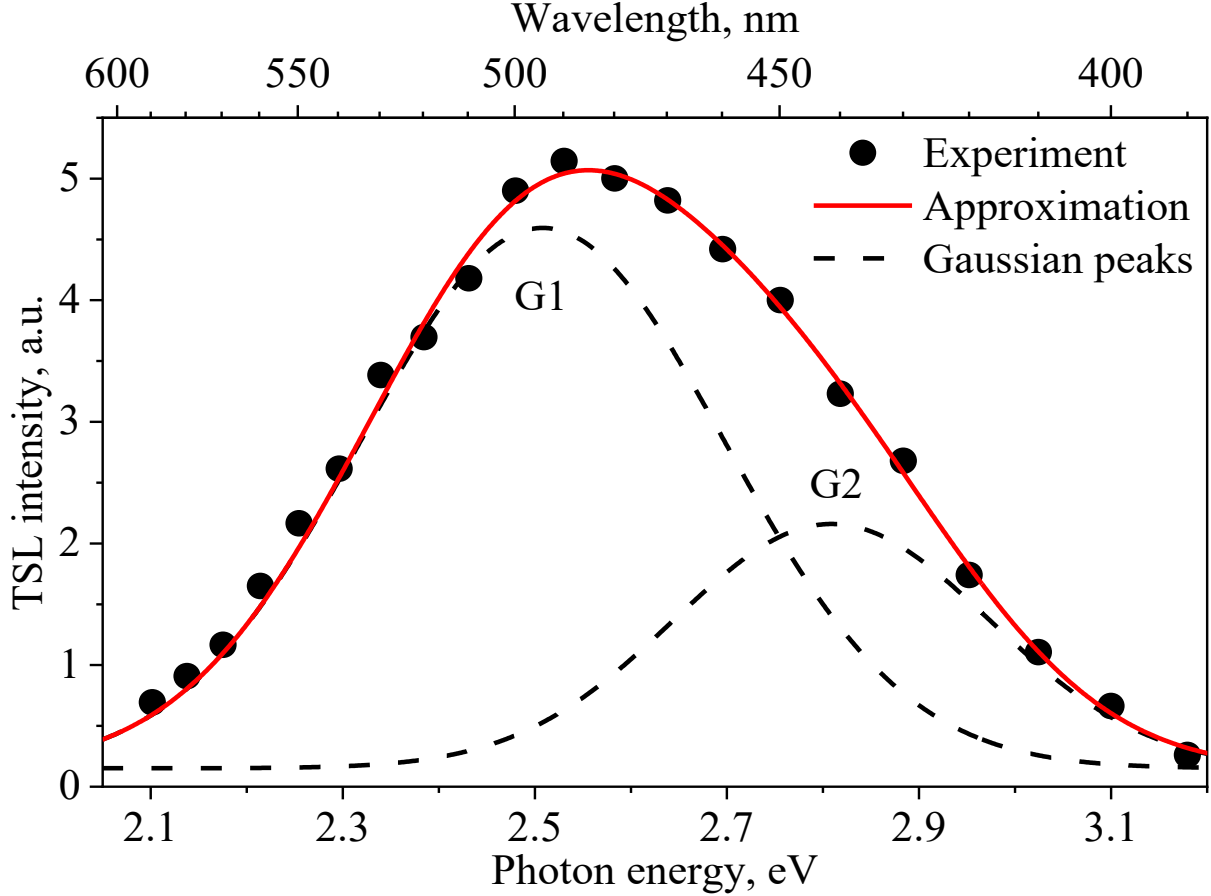

**Figure 3.** Decomposition of the TSL emission spectrum into Gaussian peaks.

*4.2. Estimates of Energy and Kinetic Parameters*

Figure 4 outlines approximations of the experimental TSL curves in the $\lambda_{em}$ = 520 and 420 nm bands that meet the emissions in the G1 and G2 regions, respectively. Numerical analysis was performed through a superposition of 4 peaks (P1–P4), according to Eq. (1). The obtained values of the calculated parameters are listed in Table 1. The coefficients of determination $R^2 > 0.997$ indicate a high matching accuracy between the predicted and experimental data.

**Table 1.** Calculated parameters of TSL processes in the 520 (G1) and 420 nm (G2) bands

| Peak | $\lambda_{em}$, nm | $T_{max}$, ± 2 K | $E_a$, ± 0.03 eV | $s''$, s$^{-1}$ | $b$, ± 0.1 |
|---|---|---|---|---|---|
| P1 | 520 | 331 | 0.67 | 2.4·10$^9$ | 1.7 |
|    | 420 | 335 | 0.68 | 1.9·10$^9$ | 1.7 |
| P2 | 520 | 363 | 0.70 | 5.1·10$^8$ | 2.0 |
|    | 420 | 371 | 0.69 | 2.8·10$^8$ | 2.0 |
| P3 | 520 | 422 | 0.72 | 4.1·10$^7$ | 1.1 |
|    | 420 | 426 | 0.69 | 1.4·10$^7$ | 1.0 |
| P4 | 520 | 475 | 0.81 | 3.5·10$^7$ | 2.0 |
|    | 420 | 482 | 0.76 | 5.8·10$^6$ | 1.7 |

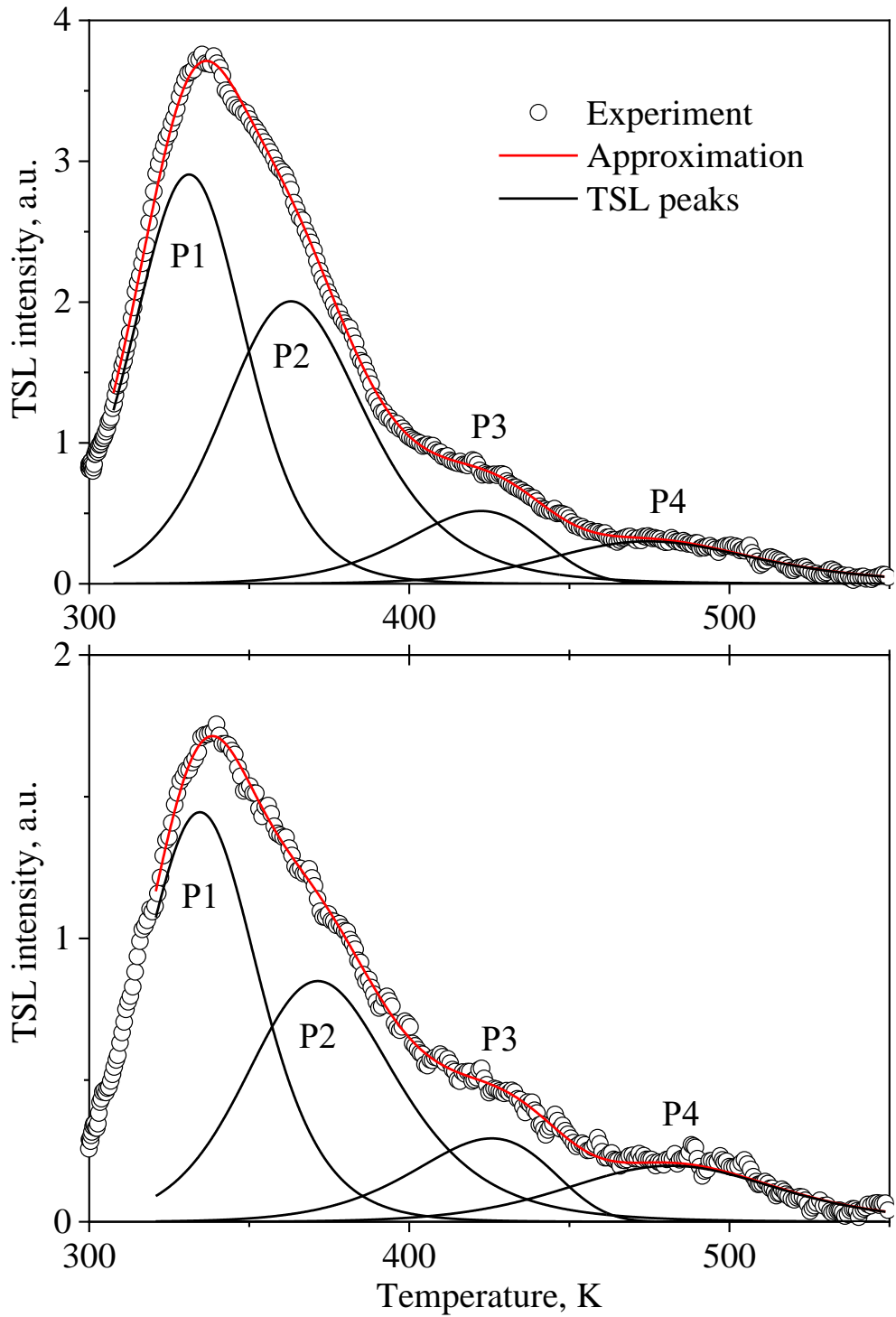

**Figure 4.** Decomposition of the TSL curve in the 520 (top) and 420 nm (bottom) bands into elementary components in the frame of the general order of kinetics by Eq. (1).

Table 1 evinces that the $T_{max}$ values for the TSL peaks in the $\lambda_{em} = 420$ nm band are shifted to lower temperatures by 4–8 K relative to those in the $\lambda_{em} = 520$ nm band. It should be emphasized that second-order kinetic processes ($b = 1.7$–$2.0$) dominate in the TSL mechanisms responsible for the P1, P2 and P4 peaks. In other words, the probability of re-capture of liberated charge carriers is extremely high. Moreover, the first-order processes ($b = 1.0$–$1.1$) proceed in the P3 peak temperature range. Simultaneously, all the traps, the depletion of which forms the observed TSL signal, are characterized by close values of the activation energy $E_a$.

The obtained values of the thermal activation process parameters (Table 1) agree satisfactorily with the TSL parameters for the monoclinic $ZrO_2$ powder in the 350–450 nm band after β-irradiation with a dose of 10 Gy [25]. In this work, the TSL curves were approximated employing four first-order kinetic peaks with close values of $E_a$ = 0.69–0.74 eV and $s$ = 3·10$^8$–9·10$^9$ s$^{-1}$. Accordingly, the authors of [25] tied the recorded TSL peaks with four different electron traps.

In our case, the values of the parameter $s''$ for the P1–P3 peaks are distinct by two orders of magnitude at close activation energies of 0.67–0.72 eV and at a change in the $T_{max}$ from 331 to 426 K for the corresponding peaks (see Table 1). Note that vacancies in atomic positions of three- and four-fold coordinated oxygen in ZrO$_2$ are electron traps with $E_a \approx$ 0.6–0.8 eV [41–46] (see Table 2). Whereas, according to calculations [47], interstitial $O_i^{2-}$, $O_i^-$ and $O_i^0$ ions surrounded by three Zr ions are hole traps and have a hole affinity energy of 0.07, 0.78 eV and 1.67 eV, respectively. The value 0.78 for $O_i^-$ ions is in agreement with $E_a$ = 0.76–0.81 eV estimated for the P4 peak (see Tables 1 and 2). The TSL spectral features revealed, the calculated parameters of trapping centers (see Table 1) and their comparative analysis with independent experimental and theoretical data [10,21–47] make it possible to propose mechanisms for the TSL processes in nanotubular zirconia arrays involving both electron and hole traps.

**Table 2.** Active traps in nanotubes ZrO$_2$

| Peak | Trap depth, eV | Trap type | Defect | Defect position | Reference |
|------|----------------|-----------|--------|-----------------|-----------|
| P1   | 0.73           | Electron  | $F^{2-}$ | Three-fold coordinated oxygen | [43] |
|      | 0.6            |           |        |                 | [44] |
| P2   | 0.76           |           | $F^-$  |                 | [43] |
|      | 0.8            |           |        | Four-fold coordinated oxygen | [46] |
| P3   | 0.81           |           | $F^{2-}$ |                 | [43] |
|      | 0.7            |           |        |                 | [44] |
|      | 0.8            |           |        |                 | [46] |
| P4   | 0.78           | Hole      | $O_i^-$ | Interstitial    | [47] |

*4.3. Mechanisms of TSL process*

Under 4.1-eV irradiation of the ZrO$_2$ nanotubes, an electron transition from the valence band (VB) into the excited level of the T-defect occurs (see Fig. 5a, ① process), followed by a non-radiative relaxation to the level of the Zr$^{3+}$-center (T-center with captured electron at Fig. 5a): T + $h\nu_{4.1eV}$ → T* → Zr$^{3+}$. The symbol * indicates the excited state of the T-center. According to [10], the effective optical bleaching of this center is oserved in a broad charge transfer band with a maximum at 3.3 eV (375 nm). Furthering absorption of a photon with an energy of 4.1 eV (see Fig. 5a, ② process) causes the electron, after being excited into the conduction band (CB), to occupy the levels of an oxygen vacancies in different charge states: $F^{2+}$ + e → $F^+$, $F^+$ + e → F, F + e → $F^-$ and $F^-$ + e → $F^{2-}$ (see Fig. 5a, ③ processes). In this case, holes emerged in the valence band are captured by the interstitial oxygen ions: $O_i^{2-}$ + h → $O_i^-$ and $O_i^-$ + h → $O_i^0$ (see Fig. 5a, ④ processes).

Then, during heating of the UV-irradiated samples, free charge carriers arise in the conduction and valence bands (see Fig. 5b, ⑤ and ⑥ processes). Subsequently, radiative recombination processes run (see Fig. 5b) involving both Zr$^{3+}$-centers: Zr$^{3+}$ + h → T + $h\nu_{2.5eV}$ (the G1 emission is formed) and $F^+$-centers: $F^+$ + e → F* → F + $h\nu_{2.8eV}$ (the G2 emission is formed), where F* corresponds to excited state of F-center. The recapture processes of freed charge carriers at the levels of electron and hole traps also occurs (see Fig. 5b, ⑦ and ⑧ processes), which is confirmed by high values of the kinetics order b = 1.7-2.0 for peaks P1, P2 and P4 $b$ = 1.7-2.0 for peaks P1, P2 и P4 (see Table 1). The Table 2 presents the origin of active traps and their corresponding positions in crystal lattice.

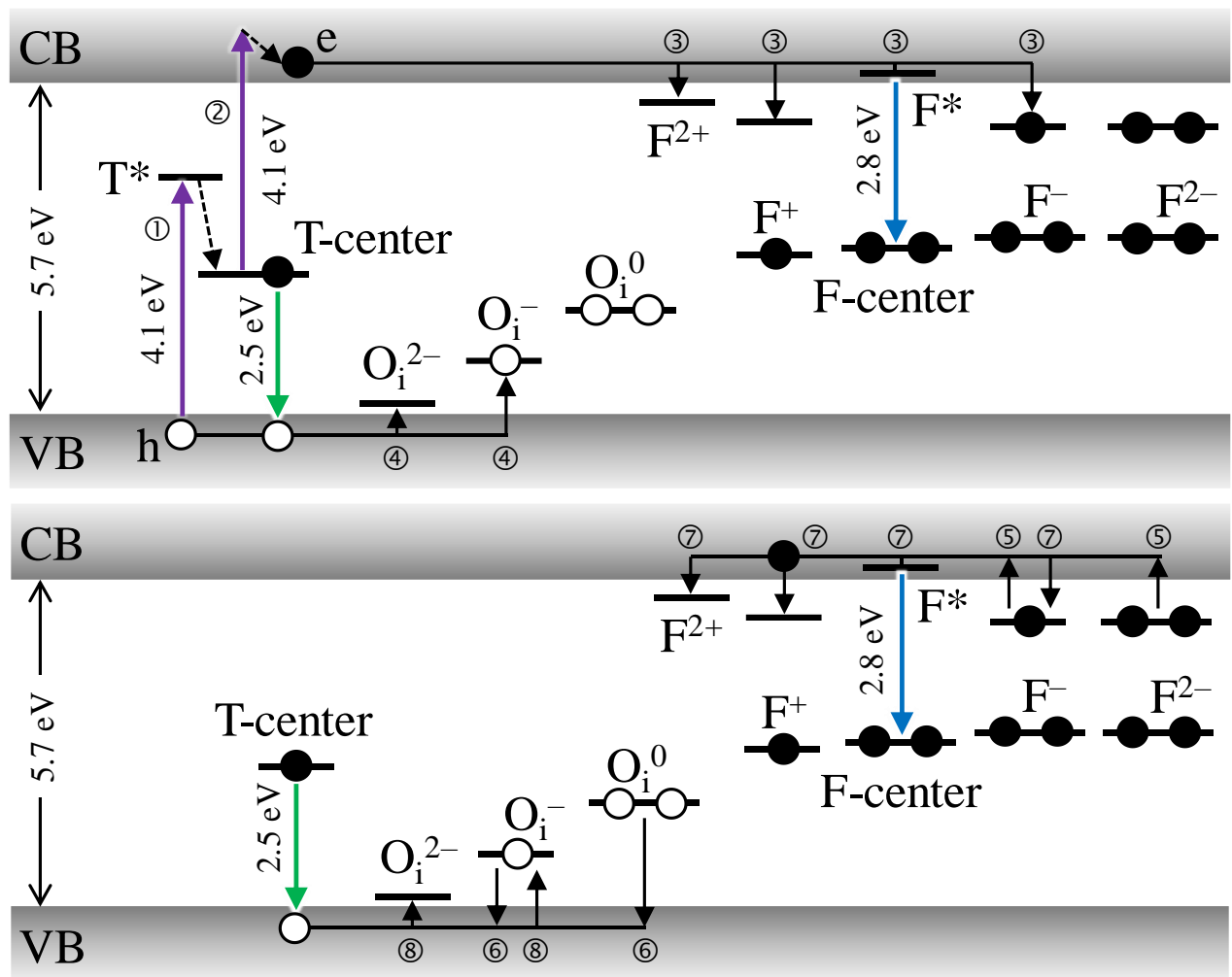

**Figure 5.** Band diagram for TSL process in ZrO$_2$ nanotubes under 4.1-eV irradiation (top) and heating (bottom) stages

**Conclusion**

Arrays, 5 μm thick, consisting of ZrO$_2$ nanotubes with an outer diameter of about 30 nm, were synthesized by anodic oxidation. Upon irradiated with monochromatic UV radiation of 4.1 eV, the ZrO$_2$ samples were studied by spectrally resolved TSL method in the range of 390–590 nm and at above-room temperatures.

It is established that the TSL emission contains two spectral components with maxima at about 2.5 and 2.8 eV. By an analysis of independent data on the luminescence of ZrO$_2$ samples, the former can be ascribed to the recombination of holes from the valence band with electrons captured by Zr$^{3+}$ ions of T-centers by the reaction Zr$^{3+}$ + h → T + $h\nu_{2.5eV}$. The latter can be attributed to the recombination processes involving F$^+$-centers by the reaction F$^+$ + e → F* → F + $h\nu_{2.8eV}$.

It is shown within the general-order kinetics, that the recorded TSL glow curves are a superposition of four temperature peaks. The energy and kinetic characteristics of the observed TSL processes are found. Also, it is shown that vacancy defects in three- and four-coordinated oxygen sites initiate a system of electron trapping levels in the ZrO$_2$ bandgap. These traps are characterized by close values of the activation energies $E_a \approx 0.7$ eV and are responsible for the emergence of three TSL peaks ranged in temperatures from 300 to 450 K. Simultaneously, a TSL peak appears in the region of > 450 K due to interstitial oxygen ions as $E_a \approx 0.8$ eV hole traps. The results obtained and the analysis of the independent data allowed one to propose the mechanisms of thermally stimulated luminescence of the subband irradiated samples with the participation of T-centers based on the bound state between two neighboring oxygen vacancies, positioned in [111] direction, with the

$Zr^{3+}$-ion. The model at hand offers the interpretation of the spectral and kinetic features of the observed thermally activated processes in UV-irradiated $ZrO_2$ nanotubes.

**Acknowledgments**

The research funding from the Ministry of Science and Higher Education of the Russian Federation (Ural Federal University project within the Priority-2030 Program) is gratefully acknowledged.